\documentclass[3p,times,sort&compress]{elsarticle}

\usepackage{ecrc}

\volume{00}

\firstpage{1}

\journalname{Nuclear Physics A}

\runauth{}

\jid{nupha}

\usepackage{amssymb}

\usepackage[figuresright]{rotating}
\usepackage{graphicx}

\newcommand{\pt}{\ensuremath{p_{T}}}
\newcommand{\kperp}{\ensuremath{k_{\perp}}}

\begin{document}

\begin{frontmatter}



\dochead{}

\title{Comparing parton energy loss models}


\author{M. van Leeuwen}

\address{Nikhef, National Institute for Subatomic Physics, P.O.Box
  41882, 1009 DB  Amsterdam, Netherlands and \\ 
  Utrecht University, P.O. Box 80000, 3508 TA  Utrecht, Netherlands}

\begin{abstract}
The similarities and differences between three commonly used
formalisms for radiative parton energy loss in hot strongly
interacting matter are discussed. The  
single gluon emission spectra are evaluated for  a model system
consisting of a homogeneous medium with a fixed length, the `TECHQM
brick'. Sizable quantitative differences are found and the origins of
these differences are discussed.
\end{abstract}


\end{frontmatter}


One of the most striking experimental results from high-energy nuclear
collisions at the Relativistic Heavy Ion Collider (RHIC) is that
hadron production at high transverse momentum \pt{} is suppressed by a factor 4--5 compared
to expectations from an independent superposition of nucleon-nucleon
collisions
\cite{Adams:2003kv,Adare:2008qa,Back:2004ra,Arsene:2003yk}. High-\pt{}
hadrons are dominantly produced by fragmentation of high-\pt{} partons
from hard scatterings in the early stage of the collision. The
suppression of hadron production is generally attributed to energy loss
of the high-momentum partons as they propagate through the hot and dense
medium.

The dominant parton energy loss mechanism is expected to be radiative
energy loss, although collisional energy loss also plays a role
\cite{Wicks:2005gt,Qin:2007rn}. These processes have been extensively
modeled with the intent to use the measured suppression to determine
medium properties like the transport coefficient or (energy) density
\cite{Wicks:2005gt,Qin:2007rn,Dainese:2004te,Renk:2006sx,Schenke:2009gb,Bass:2008rv}. It
has been found, however, that different energy loss calculations lead
to medium densities which differ by a large factor \cite{Bass:2008rv}.

There are four widely used formalisms to calculate radiative parton
energy loss in hot and dense QCD matter. Each formalism uses a
different set of approximations. In this contribution, three of the four
formalisms will be compared using a model problem: the `brick
problem', as formulated by TECHQM (Theory-Experiment Collaboration
in Hot Quark Matter). It turns out that the main quantitative
differences between the models can be attributed to specific
differences in the model assumptions and the approximations made in the
calculation. The most important differences are related to the
treatment of large angle radiation and assumptions about the
importance of interference effects between vacuum- and medium-induced radiation
and between scatterings in the medium.

\section{Energy loss formalisms}
The term radiative energy loss refers to gluon emission by a fast
quark or gluon which is stimulated by the presence of a colored
medium. It is important to realise that energetic partons also radiate
in vacuum, as part of the jet fragmentation process which 
leads to hadronisation. Medium-induced radiation is the additional
radiation that is stimulated by elastic and inelastic scatterings of
the parton in the medium.

Three of the four parton energy loss formalisms can be related to a
common path integral formalism, that was formulated by Baier,
Dokshitzer, Mueller, Peign\'e and Schiff (BDMPS)
\cite{Baier:1994bd,Baier:1996sk} and independently by Zakharov
\cite{Zakharov:1996fv}. The energy loss calculation is generally split
into two part. First, the single-emission spectrum is calculated using
the path integral or equivalent formulas. Multiple gluon
emission is calculated using an independent emission ansatz, where the
number of gluons is the integral over the single-gluon spectrum. In
the following discussion, we will focus on the single-gluon spectrum.

BDMPS evaluated the resulting energy loss in the approximation where
the medium consists of static scattering centers and the the
projectile and the outgoing gluon undergo many soft scatterings with
the medium (multiple-soft scattering approximation). Technically, this
corresponds to using the saddle point approximation of the
corresponding path integral. The original BDMPS calculation
takes a limit in which the medium length goes to infinity. Later,
Salgado and Wiedemann (SW) \cite{Wiedemann:2000tf,Salgado:2003gb} improved
the calculation to include finite-length effects. Multiple gluon
emission is calculated using a poisson ansatz.

The energy loss path integral can also be evaluated using an opacity
expansion, i.e. by organising the calculation in terms of the number of
dominant scatterings. This calculation was proposed by Gyulassy, Levai, and
Vitev (GLV) \cite{Gyulassy:1999zd} and independently by Wiedemann
\cite{Wiedemann:2000za}. In most applications, the further approximation is
made that only one scattering dominates the dynamics (single hard
scattering approximation). Multiple gluon
emission is calculated using a poisson ansatz.

Arnold, Moore, and Yaffe used a similar path integral, but evaluate
the entire problem using Hard-Thermal Loop improved finite temperature
field theory \cite{Arnold:2002ja}. In this calculation, the full
multiple scattering is taken into account and the medium is a
thermalised Quark-Gluon plasma with 
dynamical scattering centers. For this calculation, only results in
the infinite-length limit exist. Multiple gluon emission is calculated
using coupled rate equations.

The fourth commonly used parton energy loss formalism is the
Higher-Twist formalism, as formulated by Wang and Guo
\cite{Wang:2001ifa,Majumder:2009zu}. In this formalism the medium
properties enter as gluon field strength correlators into a higher
twist calculation of gluon radiation. Multiple gluon emission is
calculated using an extension of the DGLAP evolution, which is a
well-established theoretical tool for parton fragmentation.

\section{Large angle radiation}
\begin{figure}
  \includegraphics[width=0.5\textwidth]{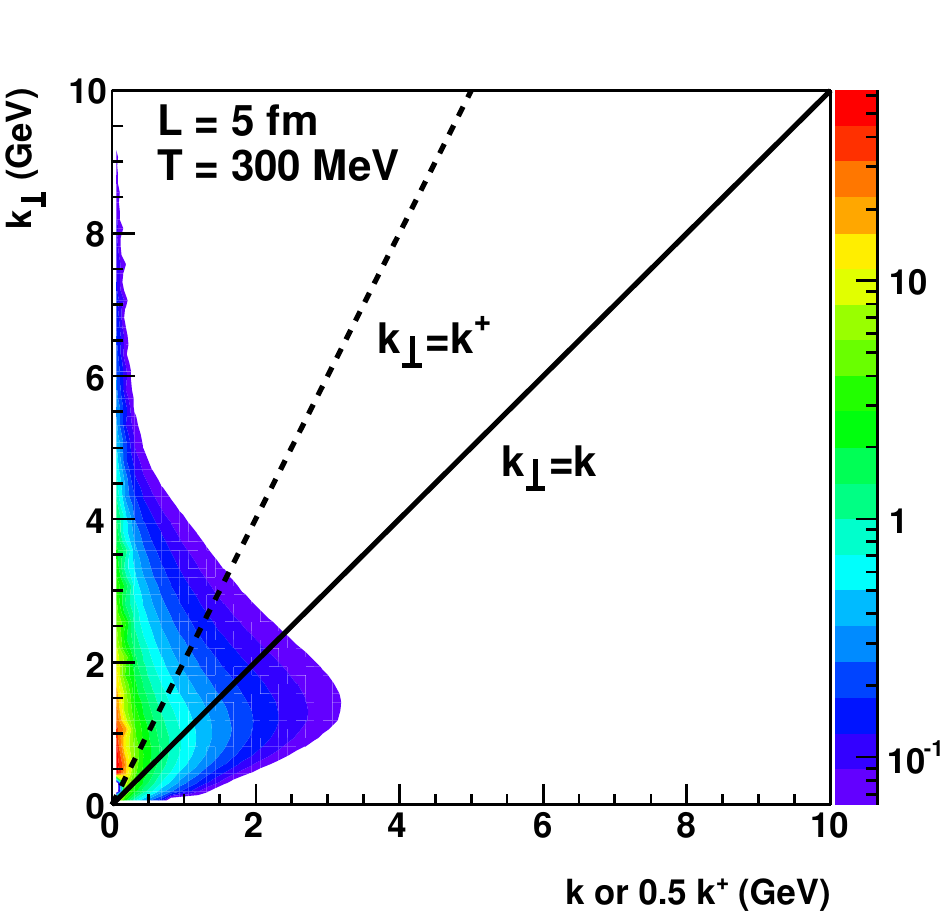}
  \includegraphics[width=0.5\textwidth]{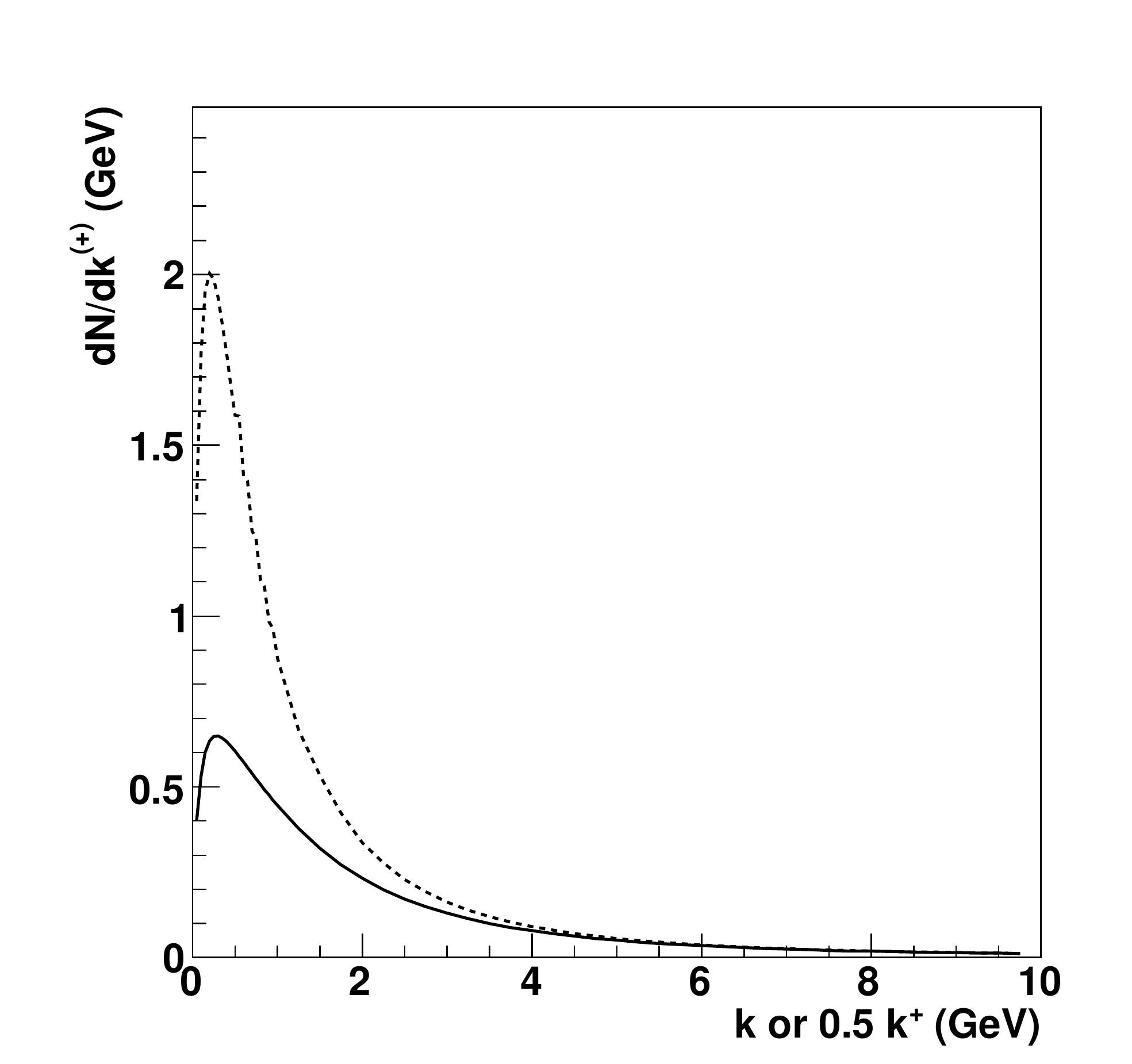}
  \caption{\label{fig:langle}Left panel: Single gluon radiation
  spectrum $dN/d\kperp dk$ from the $N=1$ opacity expansion as a
  function of total momentum $k$ or $\frac{1}{2} k^{+}$ and transverse momentum
  $k_{\perp}$. The lines indicate the kinematic limits used by SW and
  GLV (see text for details). Right panel: Projection of the gluon
  spectra on $k$ (or $\frac{1}{2}k^{+}$) using the two different kinematic limits.}
\end{figure}
All of the currently used radiative energy loss formalisms have been
derived in the soft collinear limit, in which the transverse momentum
of the radiated gluon \kperp{} is much smaller that the total gluon
momentum $k$, which in turn is much smaller than the energy $E$ of the
incoming parton $k_\perp \ll k \ll E$. The calculated gluon radiation
spectra, however, have significant radiation probabilities for large
angles, leading to important uncertainties in the calculated gluon
rates.

To illustrate this, consider the gluon radiation spectrum from the
first order opacity expansion
\begin{equation}
\label{eq:glv_rad}
x \frac{dN}{dk^2 dx} = \frac{2 C_{R} \alpha_s}{\pi} (1 - x + x^2/2)
\frac{L}{\lambda} \frac{1}{\kperp^2} \int_0^{q_{\mathrm{max}}} d^2\mathbf{q}
\frac{\mu^2}{\pi(\mathbf{q}^2 + \mu^2)^2} \times \frac{2\mathbf{k}
\cdot \mathbf{q}\;(\mathbf{k}-\mathbf{q})^2\;L^2}{16x^2E^2+(\mathbf{k}-\mathbf{q})^4\;L^2},
\end{equation}
where $L$ is the length of the medium, $\mu$ the Debye screening mass,
$E$ the parton energy, $\mathbf{k}$ the transverse momentum vector of the
emitted gluon, and $\mathbf{q}$ the transverse momentum vector exchanged with
the medium. The momentum fraction $x$ is taken to be the light-cone
fraction $x^{+}=k^{+}/E^{+}$ in the work by Gyulassy, Levai, and Vitev
\cite{Gyulassy:1999zd}, while it is taken to be a fraction in Minkowski space
$x_{E}=k/E$ by
Salgado and Wiedemann \cite{Salgado:2003gb}. Figure \ref{fig:langle} shows the
radiation spectrum Eq. \ref{eq:glv_rad} for a medium with length $L =
5$ fm and a temperature $T = 300$ MeV ($\mu^2=g^2T^2=0.34$ GeV$^2$,
$\lambda=1.0$ fm, $q_{\mathrm{max}}=\sqrt{3 \mu T}$). The emitted gluon spectrum peaks at small gluon momentum
$k$ and at finite transverse momentum $k_\perp \sim \mu$. Also shown
in the figure are lines indicating the kinematic limits (perpendicular
radiation) $\kperp = k$ and $\kperp = k^+$ in Minkowski and light-cone
coordinates. These limits are different because $x^+
\approx x_E$ and $k \approx \frac{1}{2}k^+$ only in the small-angle limit,
and the difference between $k$ and $\frac{1}{2} k^+$ becomes significant 
for large-angle radiation. 

The right panel of Fig. \ref{fig:langle} shows the gluon radiation
spectrum as a function of gluon momentum $k$ or half the light-cone
momentum $k^+$ which is obtained by integrating the
double-differential spectrum over \kperp{} within the kinematic
limits. A large difference in the radiation probability at small
momenta $k$ results from the choice of Minkowski or light-cone
coordinates. This large difference is a measure of the uncertainty
associated with applying the calculation beyond the soft-collinear
limit for which it was derived. The associated uncertainty is large
because the calculated radiation spectrum has significant intensity at
large angles. This was qualitatively pointed out already in
\cite{Wiedemann:2000tf}. For a more extensive exploration of the
differences between the opacity expansion formalism used by GLV and
SW, see \cite{Horowitz:2009eb}.

\section{AMY and the multiple soft scattering approximation}
\begin{figure}
\includegraphics[width=0.5\textwidth]{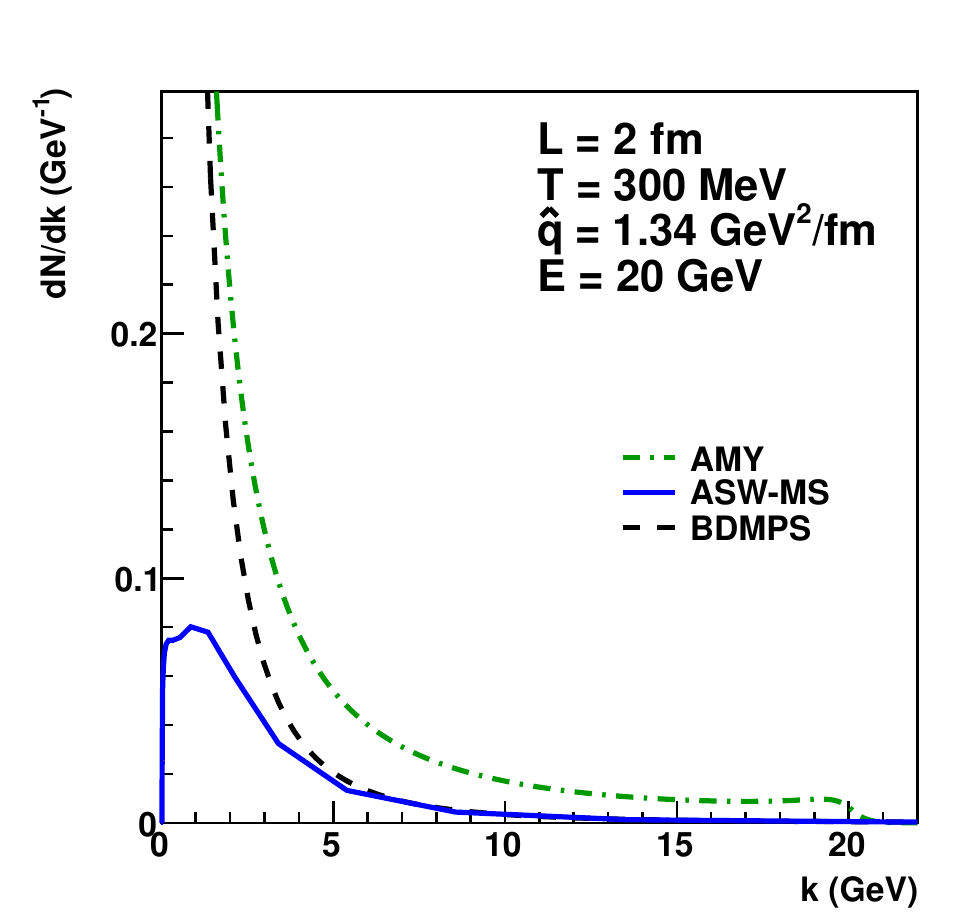}
\includegraphics[width=0.5\textwidth]{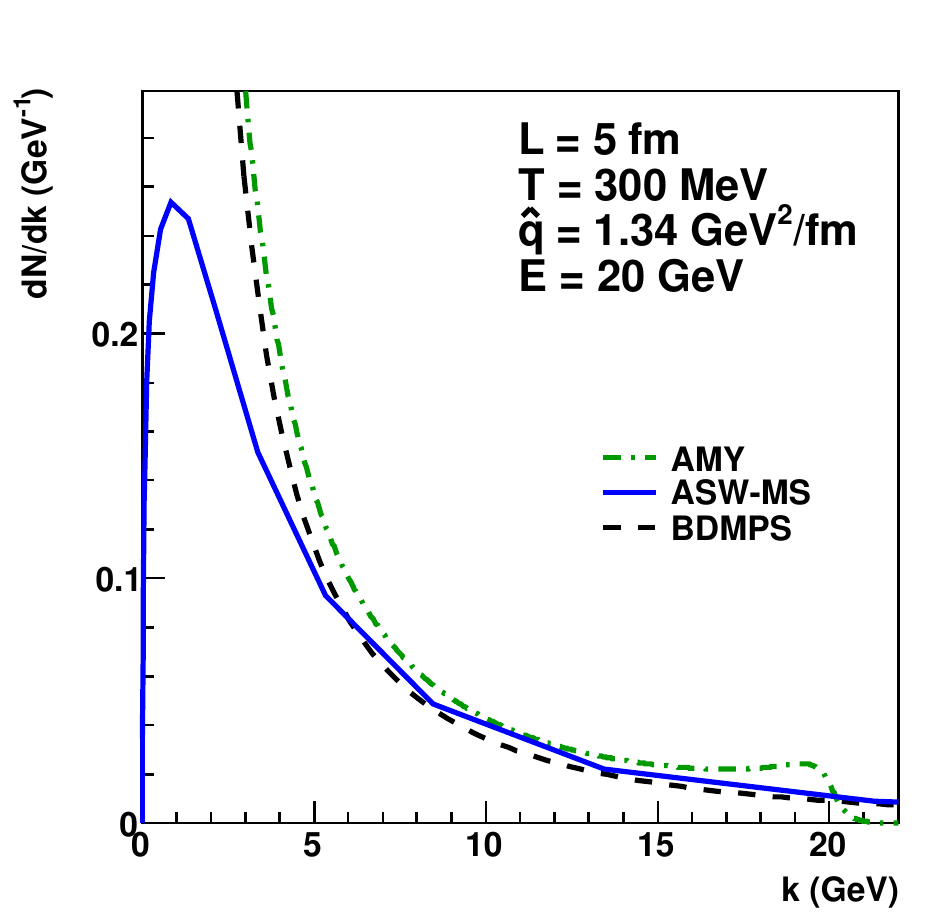}
\caption{\label{fig:AMYBDMPS}Gluon radiation spectra as a function of
  gluon momentum $k$ for AMY, ASW-MS, and the BDMPS $L\rightarrow \infty$
  limit for a medium of length $L=2$ fm (left panel) and $L=5$ fm
  (right panel).}
\end{figure}
Both the AMY formalism \cite{Arnold:2002ja} and the multiple-soft
scattering formalism formulated by BDMPS \cite{Baier:1994bd} and later
extended by SW (ASW-MS formalism)
\cite{Wiedemann:2000tf,Salgado:2003gb} start from a path-integral
formalism in which all scatterings in the medium are taken into
account. The main conceptual difference between the two calculations
is that the AMY formalism uses Hard Thermal Loop field theory, which
implies an equilibrated Quark-Gluon Plasma at high temperature as the
medium model, while BDMPS use static scattering centers
\cite{Arnold:2008iy}.

However, in addition to this conceptual difference, there are
important implementation choices which affect the gluon emission
rates. This is illustrated in Fig. \ref{fig:AMYBDMPS}, where the gluon
emission spectra for AMY, BDMPS and ASW-MS are compared for two path
lengths ($L=2$ and 5 fm). For this comparison, the transport
coefficient $\hat{q}$ which is used to set the medium density in the
BDMPS and ASW-MS calculations was calculated using the $q \ll T$ limit of
the HTL scattering rate $d\Gamma/d^2q=C_{R}/(2\pi)^2 g \mu^2
T/q^2(q^2+\mu^2)$, which is also used in the AMY calculation.

Comparing first the AMY (green dash-dotted line in Fig
\ref{fig:AMYBDMPS}) curves with the BDMPS (dashed) calculation, one
sees that the results are similar for $L=5$ fm, while for short path
lengths ($L=2$ fm), AMY generates more radiation. This is due to the
fact that AMY uses the limit $L \to \infty$ which ignores the
interference between radiation in the vacuum and the medium, and the
increased effect of finite formation times at small
length\cite{CaronHuot:2008ni}. The ASW-MS calculation (blue curve in
Fig. \ref{fig:AMYBDMPS}) is based on the BDMPS formalism, but takes
into account finite-length effects. These effects include the
interference between medium-induced and vacuum radiation as well as
the large-angle cut-off described in the previous section, although
these aspects are not easily separated in the calculation
\cite{Wiedemann:2000tf}. The effect of the finite-length corrections is
a pronounced reduction of the radiation at small gluon energies $k$,
which is similar to the effect seen in Fig. \ref{fig:langle}.

\section{Putting everything together}
\begin{figure}
\includegraphics[width=0.5\textwidth]{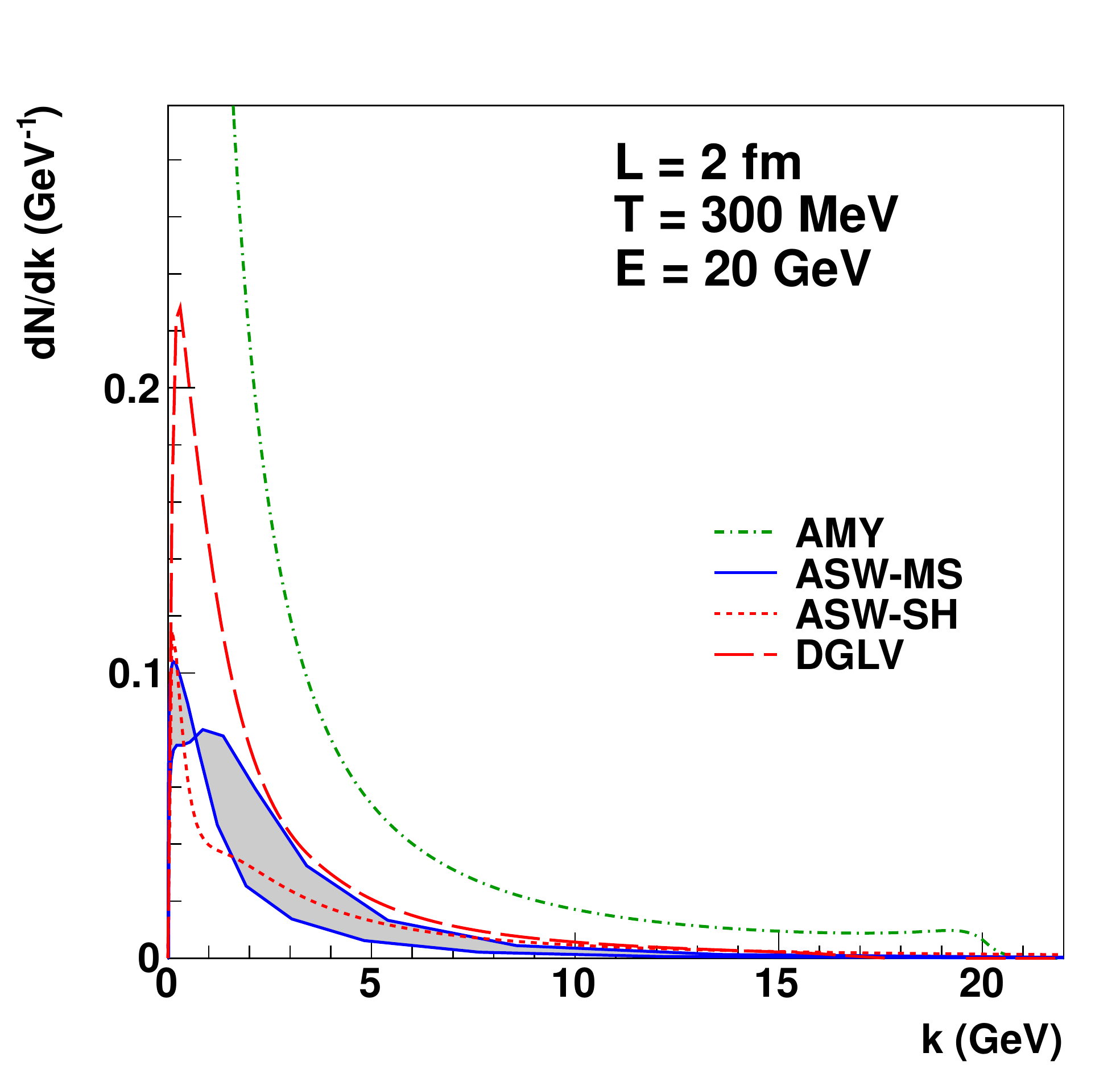}
\includegraphics[width=0.5\textwidth]{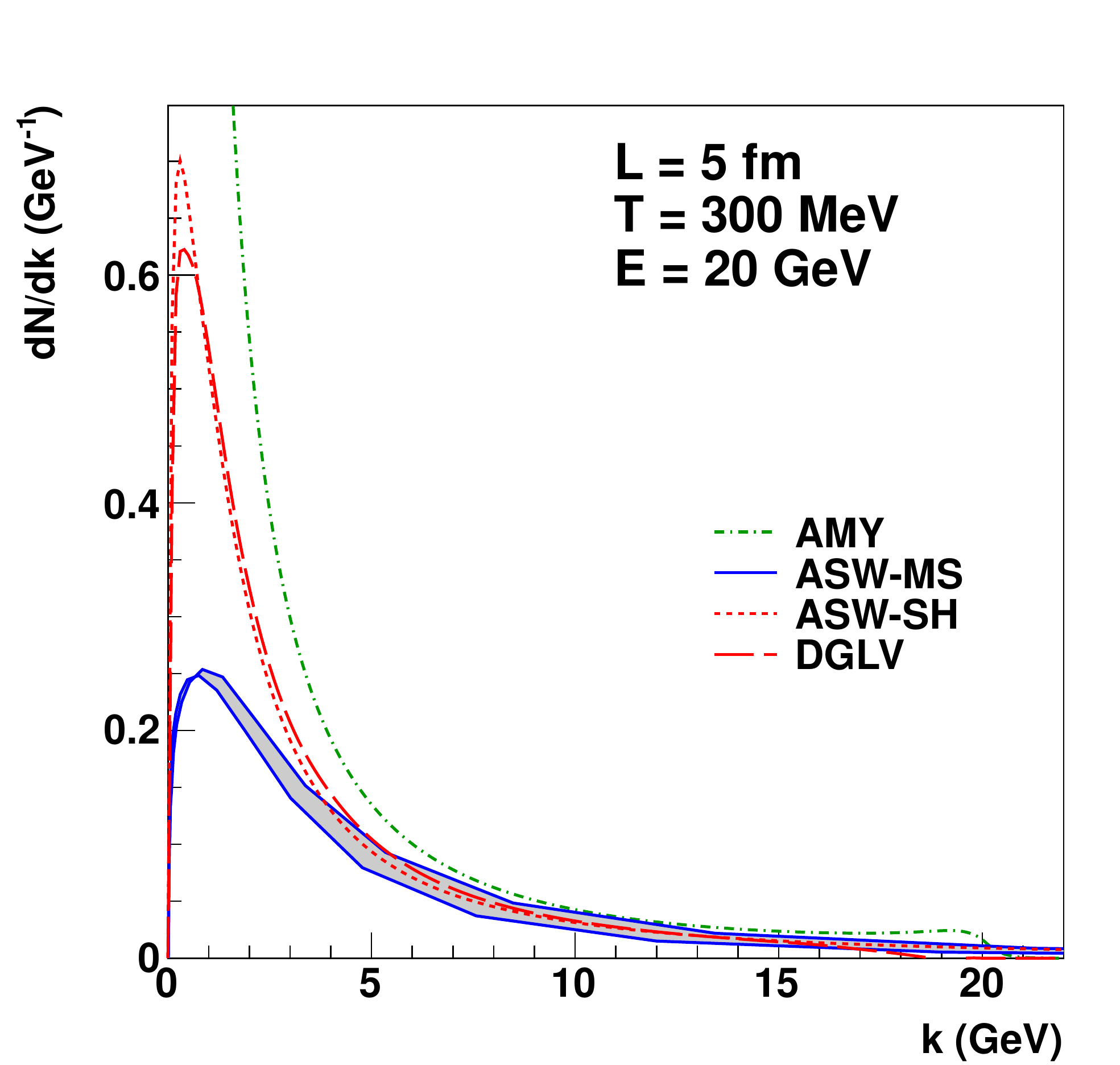}
\caption{\label{fig:compall}Gluon radiation spectrum for two variants
of the opacity expansion (DGLV and ASW-SH), the HTL-based calculation
(AMY), and the multiple-soft scattering approximation (ASW-SH). The
band for the multiple-soft scattering approximation indicates the
uncertainty associated with the calculation of the transport
coefficient $\hat{q}$ from the temperature $T$.}
\end{figure}
Figure \ref{fig:compall} shows the gluon radiation for two variants of
the opacity expansion (DGLV and ASW-SH), the HTL-based calculation
(AMY), and the multiple-soft scattering approximation (ASW-SH). The
band for the multiple-soft scattering approximation indicates the
uncertainty associated with the calculation of the transport
coefficient $\hat{q}$ from the temperature $T$. The figure clearly
shows that the AMY calculation produces the most radiation, while the
multiple-soft scattering approximation (ASW-MS) gives the least. The
results form the and the opacity expansions (ASW-SH and DGLV) are
between those extremes. The differences are
also path-length dependent.

The largest differences occur at small $k$,
where the effect of large-angle radiation is most pronounced. The AMY
calculation does not take into account the kinematic bound in this
regime, which leads to the diverging behaviour as $k \rightarrow
0$. The other calculations implement the kinematic bound, but the
associated uncertainties are large, because the emission rate near the
boundary can be large (see Fig. \ref{fig:langle}). 

A more detailed discussion of the path-length dependence of the
emission can be found in \cite{CaronHuot:2008ni}, where the different
calculations are compared to a numerical solution of the full path
integral. In that paper one can clearly see that for short path
lengths the $N=1$ opacity expansion provides the most accurate result,
while the AMY calculation overestimates the radiation and the
BDMPS/ASW-MS calculation underestimates the radiation (see also
\cite{Arnold:2009mr}). For longer path lengths, on the other hand, the
$N=1$ opacity expansion produces too much radiation, because it does
not take into account the interference between scattering centers,
while the AMY and BDMPS/ASW-MS results are in good agreement with the
path-integral formalism.

\section{Conclusion and outlook}
The side-by-side comparisons of single gluon emission spectra from
medium-induced parton energy loss using three different energy loss
formalisms as presented here clearly show sizeable differences in the
emission rates. 

Some of the differences are due to incomplete
treatment of interference effects, either between medium-induced and
vacuum radiation (AMY) or between scattering centers ($N=1$ opacity
expansion). Such approximations may be appropriate in limits of short or
long path lengths. A quantitative evaluation of the radiation process
in heavy ion collisions, however, requires calculations for short as well as long path length. Future calculations
should therefore aim to include all interference terms that are
known. It also important to include a realistic scattering potential,
that includes reasonable probabilities for large momentum transfer, which reduces the formation
time significantly \cite{Arnold:2009mr}. This is currently done in the
opacity expansions and AMY, but not in the multiple soft scattering
approximation ASW-MS.

Another important source of uncertainty in the current calculations is
the collinear approximation that is used the derivation of the radiation
spectra. All calculations, except AMY, implement a cut-off to limit
radiation to the kinematic bound at large angles ($\kperp < k$), but
the underlying formalisms generate large probability for large-angle
radiation at small momenta, thus violating the soft collinear
approximation.  This could be systematically addressed using
calculations that go beyond the collinear limit. Some work
in this direction already exists, mostly in the form of Monte-Carlo
approaches which can in principle use the full 2-to-3 matrix elements
instead of the collinear limit for in-medium scattering and radiation
\cite{Zapp:2008af,Armesto:2007dt,Armesto:2009fj}.

\section*{Acknowledgments}

The author wishes to thank the TECHQM collaboration
  for extensive and insightful discussions on the topic of parton
  energy loss and the comparison between the different formalism and
  for providing some of the curves used in Fig. \ref{fig:AMYBDMPS}
  and Fig. \ref{fig:compall}.




\bibliographystyle{model1a-num-names}
\bibliography{mvanleeuwen_eloss_HP}







\end{document}